\hsize=31pc 
\vsize=49pc 
\lineskip=0pt 
\parskip=0pt plus 1pt 
\hfuzz=1pt   
\vfuzz=2pt 
\pretolerance=2500 
\tolerance=5000 
\vbadness=5000 
\hbadness=5000 
\widowpenalty=500 
\clubpenalty=200 
\brokenpenalty=500 
\predisplaypenalty=200 
\voffset=-1pc 
\nopagenumbers      
\catcode`@=11 
\newif\ifams 
\amsfalse 
%
%
%
\newfam\bdifam 
\newfam\bsyfam 
\newfam\bssfam 
\newfam\msafam 
\newfam\msbfam 
\newif\ifxxpt    
\newif\ifxviipt  
\newif\ifxivpt   
\newif\ifxiipt   
\newif\ifxipt    
\newif\ifxpt     
\newif\ifixpt    
\newif\ifviiipt  
\newif\ifviipt   
\newif\ifvipt    
\newif\ifvpt     
%
%
\def\headsize#1#2{\def\headb@seline{#2}%
                \ifnum#1=20\def\HEAD{twenty}%
                           \def\smHEAD{twelve}%
                           \def\vsHEAD{nine}%
                           \ifxxpt\else\xdef\f@ntsize{\HEAD}%
                           \def\m@g{4}\def\s@ze{20.74}%
                           \loadheadfonts\xxpttrue\fi 
                           \ifxiipt\else\xdef\f@ntsize{\smHEAD}%
                           \def\m@g{1}\def\s@ze{12}%
                           \loadxiiptfonts\xiipttrue\fi 
                           \ifixpt\else\xdef\f@ntsize{\vsHEAD}%
                           \def\s@ze{9}%
                           \loadsmallfonts\ixpttrue\fi 
                      \else 
                \ifnum#1=17\def\HEAD{seventeen}%
                           \def\smHEAD{eleven}%
                           \def\vsHEAD{eight}%
                           \ifxviipt\else\xdef\f@ntsize{\HEAD}%
                           \def\m@g{3}\def\s@ze{17.28}%
                           \loadheadfonts\xviipttrue\fi 
                           \ifxipt\else\xdef\f@ntsize{\smHEAD}%
                           \loadxiptfonts\xipttrue\fi 
                           \ifviiipt\else\xdef\f@ntsize{\vsHEAD}%
                           \def\s@ze{8}%
                           \loadsmallfonts\viiipttrue\fi 
                      \else\def\HEAD{fourteen}%
                           \def\smHEAD{ten}%
                           \def\vsHEAD{seven}%
                           \ifxivpt\else\xdef\f@ntsize{\HEAD}%
                           \def\m@g{2}\def\s@ze{14.4}%
                           \loadheadfonts\xivpttrue\fi 
                           \ifxpt\else\xdef\f@ntsize{\smHEAD}%
                           \def\s@ze{10}%
                           \loadxptfonts\xpttrue\fi 
                           \ifviipt\else\xdef\f@ntsize{\vsHEAD}%
                           \def\s@ze{7}%
                           \loadviiptfonts\viipttrue\fi 
                \ifnum#1=14\else 
                \message{Header size should be 20, 17 or 14 point 
                              will now default to 14pt}\fi 
                \fi\fi\headfonts} 
%
%
\def\textsize#1#2{\def\textb@seline{#2}%
                 \ifnum#1=12\def\TEXT{twelve}%
                           \def\smTEXT{eight}%
                           \def\vsTEXT{six}%
                           \ifxiipt\else\xdef\f@ntsize{\TEXT}%
                           \def\m@g{1}\def\s@ze{12}%
                           \loadxiiptfonts\xiipttrue\fi 
                           \ifviiipt\else\xdef\f@ntsize{\smTEXT}%
                           \def\s@ze{8}%
                           \loadsmallfonts\viiipttrue\fi 
                           \ifvipt\else\xdef\f@ntsize{\vsTEXT}%
                           \def\s@ze{6}%
                           \loadviptfonts\vipttrue\fi 
                      \else 
                \ifnum#1=11\def\TEXT{eleven}%
                           \def\smTEXT{seven}%
                           \def\vsTEXT{five}%
                           \ifxipt\else\xdef\f@ntsize{\TEXT}%
                           \def\s@ze{11}%
                           \loadxiptfonts\xipttrue\fi 
                           \ifviipt\else\xdef\f@ntsize{\smTEXT}%
                           \loadviiptfonts\viipttrue\fi 
                           \ifvpt\else\xdef\f@ntsize{\vsTEXT}%
                           \def\s@ze{5}%
                           \loadvptfonts\vpttrue\fi 
                      \else\def\TEXT{ten}%
                           \def\smTEXT{seven}%
                           \def\vsTEXT{five}%
                           \ifxpt\else\xdef\f@ntsize{\TEXT}%
                           \loadxptfonts\xpttrue\fi 
                           \ifviipt\else\xdef\f@ntsize{\smTEXT}%
                           \def\s@ze{7}%
                           \loadviiptfonts\viipttrue\fi 
                           \ifvpt\else\xdef\f@ntsize{\vsTEXT}%
                           \def\s@ze{5}%
                           \loadvptfonts\vpttrue\fi 
                \ifnum#1=10\else 
                \message{Text size should be 12, 11 or 10 point 
                              will now default to 10pt}\fi 
                \fi\fi\textfonts} 
%
%
\def\smallsize#1#2{\def\smallb@seline{#2}%
                 \ifnum#1=10\def\SMALL{ten}%
                           \def\smSMALL{seven}%
                           \def\vsSMALL{five}%
                           \ifxpt\else\xdef\f@ntsize{\SMALL}%
                           \loadxptfonts\xpttrue\fi 
                           \ifviipt\else\xdef\f@ntsize{\smSMALL}%
                           \def\s@ze{7}%
                           \loadviiptfonts\viipttrue\fi 
                           \ifvpt\else\xdef\f@ntsize{\vsSMALL}%
                           \def\s@ze{5}%
                           \loadvptfonts\vpttrue\fi 
                       \else 
                 \ifnum#1=9\def\SMALL{nine}%
                           \def\smSMALL{six}%
                           \def\vsSMALL{five}%
                           \ifixpt\else\xdef\f@ntsize{\SMALL}%
                           \def\s@ze{9}%
                           \loadsmallfonts\ixpttrue\fi 
                           \ifvipt\else\xdef\f@ntsize{\smSMALL}%
                           \def\s@ze{6}%
                           \loadviptfonts\vipttrue\fi 
                           \ifvpt\else\xdef\f@ntsize{\vsSMALL}%
                           \def\s@ze{5}%
                           \loadvptfonts\vpttrue\fi 
                       \else 
                           \def\SMALL{eight}%
                           \def\smSMALL{six}%
                           \def\vsSMALL{five}%
                           \ifviiipt\else\xdef\f@ntsize{\SMALL}%
                           \def\s@ze{8}%
                           \loadsmallfonts\viiipttrue\fi 
                           \ifvipt\else\xdef\f@ntsize{\smSMALL}%
                           \def\s@ze{6}%
                           \loadviptfonts\vipttrue\fi 
                           \ifvpt\else\xdef\f@ntsize{\vsSMALL}%
                           \def\s@ze{5}%
                           \loadvptfonts\vpttrue\fi 
                 \ifnum#1=8\else\message{Small size should be 10, 9 or  
                            8 point will now default to 8pt}\fi 
                \fi\fi\smallfonts} 
\def\F@nt{\expandafter\font\csname} 
\def\Sk@w{\expandafter\skewchar\csname} 
\def\@nd{\endcsname} 
\def\@step#1{ scaled \magstep#1} 
\def\@half{ scaled \magstephalf} 
\def\@t#1{ at #1pt} 
%
%
\def\loadheadfonts{\bigf@nts 
\F@nt \f@ntsize bdi\@nd=cmmib10 \@t{\s@ze}%
\Sk@w \f@ntsize bdi\@nd='177 
\F@nt \f@ntsize bsy\@nd=cmbsy10 \@t{\s@ze}%
\Sk@w \f@ntsize bsy\@nd='60 
\F@nt \f@ntsize bss\@nd=cmssbx10 \@t{\s@ze}} 
%
%
\def\loadxiiptfonts{\bigf@nts 
\F@nt \f@ntsize bdi\@nd=cmmib10 \@step{\m@g}%
\Sk@w \f@ntsize bdi\@nd='177 
\F@nt \f@ntsize bsy\@nd=cmbsy10 \@step{\m@g}%
\Sk@w \f@ntsize bsy\@nd='60 
\F@nt \f@ntsize bss\@nd=cmssbx10 \@step{\m@g}} 
%
%
\def\loadxiptfonts{%
\font\elevenrm=cmr10 \@half 
\font\eleveni=cmmi10 \@half 
\skewchar\eleveni='177 
\font\elevensy=cmsy10 \@half 
\skewchar\elevensy='60 
\font\elevenex=cmex10 \@half 
\font\elevenit=cmti10 \@half 
\font\elevensl=cmsl10 \@half 
\font\elevenbf=cmbx10 \@half 
\font\eleventt=cmtt10 \@half 
\ifams\font\elevenmsa=msam10 \@half 
\font\elevenmsb=msbm10 \@half\else\fi 
\font\elevenbdi=cmmib10 \@half 
\skewchar\elevenbdi='177 
\font\elevenbsy=cmbsy10 \@half 
\skewchar\elevenbsy='60 
\font\elevenbss=cmssbx10 \@half} 
%
%
\def\loadxptfonts{%
\font\tenbdi=cmmib10 
\skewchar\tenbdi='177 
\font\tenbsy=cmbsy10  
\skewchar\tenbsy='60 
\ifams\font\tenmsa=msam10  
\font\tenmsb=msbm10\else\fi 
\font\tenbss=cmssbx10}%
%
%
\def\loadsmallfonts{\smallf@nts 
\ifams 
\F@nt \f@ntsize ex\@nd=cmex\s@ze 
\else 
\F@nt \f@ntsize ex\@nd=cmex10\fi 
\F@nt \f@ntsize it\@nd=cmti\s@ze 
\F@nt \f@ntsize sl\@nd=cmsl\s@ze 
\F@nt \f@ntsize tt\@nd=cmtt\s@ze} 
%
%
\def\loadviiptfonts{%
\font\sevenit=cmti7 
\font\sevensl=cmsl8 at 7pt 
\ifams\font\sevenmsa=msam7  
\font\sevenmsb=msbm7 
\font\sevenex=cmex7 
\font\sevenbsy=cmbsy7 
\font\sevenbdi=cmmib7\else 
\font\sevenex=cmex10 
\font\sevenbsy=cmbsy10 at 7pt 
\font\sevenbdi=cmmib10 at 7pt\fi 
\skewchar\sevenbsy='60 
\skewchar\sevenbdi='177 
\font\sevenbss=cmssbx10 at 7pt}%
%
%
\def\loadviptfonts{\smallf@nts 
\ifams\font\sixex=cmex7 at 6pt\else 
\font\sixex=cmex10\fi 
\font\sixit=cmti7 at 6pt} 
%
%
\def\loadvptfonts{%
\font\fiveit=cmti7 at 5pt 
\ifams\font\fiveex=cmex7 at 5pt 
\font\fivebdi=cmmib5 
\font\fivebsy=cmbsy5 
\font\fivemsa=msam5  
\font\fivemsb=msbm5\else 
\font\fiveex=cmex10 
\font\fivebdi=cmmib10 at 5pt 
\font\fivebsy=cmbsy10 at 5pt\fi 
\skewchar\fivebdi='177 
\skewchar\fivebsy='60 
\font\fivebss=cmssbx10 at 5pt} 
\def\bigf@nts{%
\F@nt \f@ntsize rm\@nd=cmr10 \@step{\m@g}%
\F@nt \f@ntsize i\@nd=cmmi10 \@step{\m@g}%
\Sk@w \f@ntsize i\@nd='177 
\F@nt \f@ntsize sy\@nd=cmsy10 \@step{\m@g}%
\Sk@w \f@ntsize sy\@nd='60 
\F@nt \f@ntsize ex\@nd=cmex10 \@step{\m@g}%
\F@nt \f@ntsize it\@nd=cmti10 \@step{\m@g}%
\F@nt \f@ntsize sl\@nd=cmsl10 \@step{\m@g}%
\F@nt \f@ntsize bf\@nd=cmbx10 \@step{\m@g}%
\F@nt \f@ntsize tt\@nd=cmtt10 \@step{\m@g}%
\ifams 
\F@nt \f@ntsize msa\@nd=msam10 \@step{\m@g}%
\F@nt \f@ntsize msb\@nd=msbm10 \@step{\m@g}\else\fi} 
\def\smallf@nts{%
\F@nt \f@ntsize rm\@nd=cmr\s@ze 
\F@nt \f@ntsize i\@nd=cmmi\s@ze  
\Sk@w \f@ntsize i\@nd='177 
\F@nt \f@ntsize sy\@nd=cmsy\s@ze 
\Sk@w \f@ntsize sy\@nd='60 
\F@nt \f@ntsize bf\@nd=cmbx\s@ze  
\ifams 
\F@nt \f@ntsize bdi\@nd=cmmib\s@ze  
\F@nt \f@ntsize bsy\@nd=cmbsy\s@ze  
\F@nt \f@ntsize msa\@nd=msam\s@ze  
\F@nt \f@ntsize msb\@nd=msbm\s@ze 
\else 
\F@nt \f@ntsize bdi\@nd=cmmib10 \@t{\s@ze}%
\F@nt \f@ntsize bsy\@nd=cmbsy10 \@t{\s@ze}\fi  
\Sk@w \f@ntsize bdi\@nd='177 
\Sk@w \f@ntsize bsy\@nd='60 
\F@nt \f@ntsize bss\@nd=cmssbx10 \@t{\s@ze}}%
%
%
\def\headfonts{%
\textfont0=\csname\HEAD rm\@nd         
\scriptfont0=\csname\smHEAD rm\@nd 
\scriptscriptfont0=\csname\vsHEAD rm\@nd 
\def\rm{\fam0\csname\HEAD rm\@nd 
\def\sc{\csname\smHEAD rm\@nd}}%
\textfont1=\csname\HEAD i\@nd          
\scriptfont1=\csname\smHEAD i\@nd 
\scriptscriptfont1=\csname\vsHEAD i\@nd 
\textfont2=\csname\HEAD sy\@nd         
\scriptfont2=\csname\smHEAD sy\@nd 
\scriptscriptfont2=\csname\vsHEAD sy\@nd 
\textfont3=\csname\HEAD ex\@nd         
\scriptfont3=\csname\smHEAD ex\@nd 
\scriptscriptfont3=\csname\smHEAD ex\@nd 
\textfont\itfam=\csname\HEAD it\@nd    
\scriptfont\itfam=\csname\smHEAD it\@nd 
\scriptscriptfont\itfam=\csname\vsHEAD it\@nd 
\def\it{\fam\itfam\csname\HEAD it\@nd 
\def\sc{\csname\smHEAD it\@nd}}%
\textfont\slfam=\csname\HEAD sl\@nd    
\def\sl{\fam\slfam\csname\HEAD sl\@nd 
\def\sc{\csname\smHEAD sl\@nd}}%
\textfont\bffam=\csname\HEAD bf\@nd    
\scriptfont\bffam=\csname\smHEAD bf\@nd 
\scriptscriptfont\bffam=\csname\vsHEAD bf\@nd 
\def\bf{\fam\bffam\csname\HEAD bf\@nd 
\def\sc{\csname\smHEAD bf\@nd}}%
\textfont\ttfam=\csname\HEAD tt\@nd    
\def\tt{\fam\ttfam\csname\HEAD tt\@nd}%
\textfont\bdifam=\csname\HEAD bdi\@nd  
\scriptfont\bdifam=\csname\smHEAD bdi\@nd 
\scriptscriptfont\bdifam=\csname\vsHEAD bdi\@nd 
\def\bdi{\fam\bdifam\csname\HEAD bdi\@nd}%
\textfont\bsyfam=\csname\HEAD bsy\@nd  
\scriptfont\bsyfam=\csname\smHEAD bsy\@nd 
\def\bsy{\fam\bsyfam\csname\HEAD bsy\@nd}%
\textfont\bssfam=\csname\HEAD bss\@nd  
\scriptfont\bssfam=\csname\smHEAD bss\@nd 
\scriptscriptfont\bssfam=\csname\vsHEAD bss\@nd 
\def\bss{\fam\bssfam\csname\HEAD bss\@nd}%
\ifams 
\textfont\msafam=\csname\HEAD msa\@nd  
\scriptfont\msafam=\csname\smHEAD msa\@nd 
\scriptscriptfont\msafam=\csname\vsHEAD msa\@nd 
\textfont\msbfam=\csname\HEAD msb\@nd  
\scriptfont\msbfam=\csname\smHEAD msb\@nd 
\scriptscriptfont\msbfam=\csname\vsHEAD msb\@nd 
\else\fi 
\normalbaselineskip=\headb@seline pt%
\setbox\strutbox=\hbox{\vrule height.7\normalbaselineskip  
depth.3\baselineskip width0pt}%
\def\sc{\csname\smHEAD rm\@nd}\normalbaselines\bf} 
%
%
\def\textfonts{%
\textfont0=\csname\TEXT rm\@nd         
\scriptfont0=\csname\smTEXT rm\@nd 
\scriptscriptfont0=\csname\vsTEXT rm\@nd 
\def\rm{\fam0\csname\TEXT rm\@nd 
\def\sc{\csname\smTEXT rm\@nd}}%
\textfont1=\csname\TEXT i\@nd          
\scriptfont1=\csname\smTEXT i\@nd 
\scriptscriptfont1=\csname\vsTEXT i\@nd 
\textfont2=\csname\TEXT sy\@nd         
\scriptfont2=\csname\smTEXT sy\@nd 
\scriptscriptfont2=\csname\vsTEXT sy\@nd 
\textfont3=\csname\TEXT ex\@nd         
\scriptfont3=\csname\smTEXT ex\@nd 
\scriptscriptfont3=\csname\smTEXT ex\@nd 
\textfont\itfam=\csname\TEXT it\@nd    
\scriptfont\itfam=\csname\smTEXT it\@nd 
\scriptscriptfont\itfam=\csname\vsTEXT it\@nd 
\def\it{\fam\itfam\csname\TEXT it\@nd 
\def\sc{\csname\smTEXT it\@nd}}%
\textfont\slfam=\csname\TEXT sl\@nd    
\def\sl{\fam\slfam\csname\TEXT sl\@nd 
\def\sc{\csname\smTEXT sl\@nd}}%
\textfont\bffam=\csname\TEXT bf\@nd    
\scriptfont\bffam=\csname\smTEXT bf\@nd 
\scriptscriptfont\bffam=\csname\vsTEXT bf\@nd 
\def\bf{\fam\bffam\csname\TEXT bf\@nd 
\def\sc{\csname\smTEXT bf\@nd}}%
\textfont\ttfam=\csname\TEXT tt\@nd    
\def\tt{\fam\ttfam\csname\TEXT tt\@nd}%
\textfont\bdifam=\csname\TEXT bdi\@nd  
\scriptfont\bdifam=\csname\smTEXT bdi\@nd 
\scriptscriptfont\bdifam=\csname\vsTEXT bdi\@nd 
\def\bdi{\fam\bdifam\csname\TEXT bdi\@nd}%
\textfont\bsyfam=\csname\TEXT bsy\@nd  
\scriptfont\bsyfam=\csname\smTEXT bsy\@nd 
\def\bsy{\fam\bsyfam\csname\TEXT bsy\@nd}%
\textfont\bssfam=\csname\TEXT bss\@nd  
\scriptfont\bssfam=\csname\smTEXT bss\@nd 
\scriptscriptfont\bssfam=\csname\vsTEXT bss\@nd 
\def\bss{\fam\bssfam\csname\TEXT bss\@nd}%
\ifams 
\textfont\msafam=\csname\TEXT msa\@nd  
\scriptfont\msafam=\csname\smTEXT msa\@nd 
\scriptscriptfont\msafam=\csname\vsTEXT msa\@nd 
\textfont\msbfam=\csname\TEXT msb\@nd  
\scriptfont\msbfam=\csname\smTEXT msb\@nd 
\scriptscriptfont\msbfam=\csname\vsTEXT msb\@nd 
\else\fi 
\normalbaselineskip=\textb@seline pt 
\setbox\strutbox=\hbox{\vrule height.7\normalbaselineskip  
depth.3\baselineskip width0pt}%
\everymath{}%
\def\sc{\csname\smTEXT rm\@nd}\normalbaselines\rm} 
%
%
\def\smallfonts{%
\textfont0=\csname\SMALL rm\@nd         
\scriptfont0=\csname\smSMALL rm\@nd 
\scriptscriptfont0=\csname\vsSMALL rm\@nd 
\def\rm{\fam0\csname\SMALL rm\@nd 
\def\sc{\csname\smSMALL rm\@nd}}%
\textfont1=\csname\SMALL i\@nd          
\scriptfont1=\csname\smSMALL i\@nd 
\scriptscriptfont1=\csname\vsSMALL i\@nd 
\textfont2=\csname\SMALL sy\@nd         
\scriptfont2=\csname\smSMALL sy\@nd 
\scriptscriptfont2=\csname\vsSMALL sy\@nd 
\textfont3=\csname\SMALL ex\@nd         
\scriptfont3=\csname\smSMALL ex\@nd 
\scriptscriptfont3=\csname\smSMALL ex\@nd 
\textfont\itfam=\csname\SMALL it\@nd    
\scriptfont\itfam=\csname\smSMALL it\@nd 
\scriptscriptfont\itfam=\csname\vsSMALL it\@nd 
\def\it{\fam\itfam\csname\SMALL it\@nd 
\def\sc{\csname\smSMALL it\@nd}}%
\textfont\slfam=\csname\SMALL sl\@nd    
\def\sl{\fam\slfam\csname\SMALL sl\@nd 
\def\sc{\csname\smSMALL sl\@nd}}%
\textfont\bffam=\csname\SMALL bf\@nd    
\scriptfont\bffam=\csname\smSMALL bf\@nd 
\scriptscriptfont\bffam=\csname\vsSMALL bf\@nd 
\def\bf{\fam\bffam\csname\SMALL bf\@nd 
\def\sc{\csname\smSMALL bf\@nd}}%
\textfont\ttfam=\csname\SMALL tt\@nd    
\def\tt{\fam\ttfam\csname\SMALL tt\@nd}%
\textfont\bdifam=\csname\SMALL bdi\@nd  
\scriptfont\bdifam=\csname\smSMALL bdi\@nd 
\scriptscriptfont\bdifam=\csname\vsSMALL bdi\@nd 
\def\bdi{\fam\bdifam\csname\SMALL bdi\@nd}%
\textfont\bsyfam=\csname\SMALL bsy\@nd  
\scriptfont\bsyfam=\csname\smSMALL bsy\@nd 
\def\bsy{\fam\bsyfam\csname\SMALL bsy\@nd}%
\textfont\bssfam=\csname\SMALL bss\@nd  
\scriptfont\bssfam=\csname\smSMALL bss\@nd 
\scriptscriptfont\bssfam=\csname\vsSMALL bss\@nd 
\def\bss{\fam\bssfam\csname\SMALL bss\@nd}%
\ifams 
\textfont\msafam=\csname\SMALL msa\@nd  
\scriptfont\msafam=\csname\smSMALL msa\@nd 
\scriptscriptfont\msafam=\csname\vsSMALL msa\@nd 
\textfont\msbfam=\csname\SMALL msb\@nd  
\scriptfont\msbfam=\csname\smSMALL msb\@nd 
\scriptscriptfont\msbfam=\csname\vsSMALL msb\@nd 
\else\fi 
\normalbaselineskip=\smallb@seline pt%
\setbox\strutbox=\hbox{\vrule height.7\normalbaselineskip  
depth.3\baselineskip width0pt}%
\everymath{}%
\def\sc{\csname\smSMALL rm\@nd}\normalbaselines\rm}%
\everydisplay{\indenteddisplay 
   \gdef\labeltype{\eqlabel}}%
%
%
\def\hexnumber@#1{\ifcase#1 0\or 1\or 2\or 3\or 4\or 5\or 6\or 7\or 8\or 
 9\or A\or B\or C\or D\or E\or F\fi} 
\edef\bffam@{\hexnumber@\bffam} 
\edef\bdifam@{\hexnumber@\bdifam} 
\edef\bsyfam@{\hexnumber@\bsyfam} 
\def\undefine#1{\let#1\undefined} 
\def\newsymbol#1#2#3#4#5{\let\next@\relax 
 \ifnum#2=\thr@@\let\next@\bdifam@\else 
 \ifams 
 \ifnum#2=\@ne\let\next@\msafam@\else 
 \ifnum#2=\tw@\let\next@\msbfam@\fi\fi 
 \fi\fi 
 \mathchardef#1="#3\next@#4#5} 
\def\mathhexbox@#1#2#3{\relax 
 \ifmmode\mathpalette{}{\m@th\mathchar"#1#2#3}%
 \else\leavevmode\hbox{$\m@th\mathchar"#1#2#3$}\fi} 

\def\bi#1{{\fam\bdifam\relax#1}} 
%
%
\ifams\input amsmacro\fi 
%
%
\newsymbol\bitGamma 3000 
\newsymbol\bitDelta 3001 
\newsymbol\bitTheta 3002 
\newsymbol\bitLambda 3003 
\newsymbol\bitXi 3004 
\newsymbol\bitPi 3005 
\newsymbol\bitSigma 3006 
\newsymbol\bitUpsilon 3007 
\newsymbol\bitPhi 3008 
\newsymbol\bitPsi 3009 
\newsymbol\bitOmega 300A 
\newsymbol\balpha 300B 
\newsymbol\bbeta 300C 
\newsymbol\bgamma 300D 
\newsymbol\bdelta 300E 
\newsymbol\bepsilon 300F 
\newsymbol\bzeta 3010 
\newsymbol\bfeta 3011 
\newsymbol\btheta 3012 
\newsymbol\biota 3013 
\newsymbol\bkappa 3014 
\newsymbol\blambda 3015 
\newsymbol\bmu 3016 
\newsymbol\bnu 3017 
\newsymbol\bxi 3018 
\newsymbol\bpi 3019 
\newsymbol\brho 301A 
\newsymbol\bsigma 301B 
\newsymbol\btau 301C 
\newsymbol\bupsilon 301D 
\newsymbol\bphi 301E 
\newsymbol\bchi 301F 
\newsymbol\bpsi 3020 
\newsymbol\bomega 3021 
\newsymbol\bvarepsilon 3022 
\newsymbol\bvartheta 3023 
\newsymbol\bvaromega 3024 
\newsymbol\bvarrho 3025 
\newsymbol\bvarzeta 3026 
\newsymbol\bvarphi 3027 
\newsymbol\bpartial 3040 
\newsymbol\bell 3060 
\newsymbol\bimath 307B 
\newsymbol\bjmath 307C 
\mathchardef\binfty "0\bsyfam@31 
\mathchardef\bnabla "0\bsyfam@72 
\mathchardef\bdot "2\bsyfam@01 
\mathchardef\bGamma "0\bffam@00 
\mathchardef\bDelta "0\bffam@01 
\mathchardef\bTheta "0\bffam@02 
\mathchardef\bLambda "0\bffam@03 
\mathchardef\bXi "0\bffam@04 
\mathchardef\bPi "0\bffam@05 
\mathchardef\bSigma "0\bffam@06 
\mathchardef\bUpsilon "0\bffam@07 
\mathchardef\bPhi "0\bffam@08 
\mathchardef\bPsi "0\bffam@09 
\mathchardef\bOmega "0\bffam@0A 
\mathchardef\itGamma "0100 
\mathchardef\itDelta "0101 
\mathchardef\itTheta "0102 
\mathchardef\itLambda "0103 
\mathchardef\itXi "0104 
\mathchardef\itPi "0105 
\mathchardef\itSigma "0106 
\mathchardef\itUpsilon "0107 
\mathchardef\itPhi "0108 
\mathchardef\itPsi "0109 
\mathchardef\itOmega "010A 
\mathchardef\Gamma "0000 
\mathchardef\Delta "0001 
\mathchardef\Theta "0002 
\mathchardef\Lambda "0003 
\mathchardef\Xi "0004 
\mathchardef\Pi "0005 
\mathchardef\Sigma "0006 
\mathchardef\Upsilon "0007 
\mathchardef\Phi "0008 
\mathchardef\Psi "0009 
\mathchardef\Omega "000A 
%
%
\newcount\firstpage  \firstpage=1  
\newcount\jnl                      
\newcount\secno                    
\newcount\subno                    
\newcount\subsubno                 
\newcount\appno                    
\newcount\tabno                    
\newcount\figno                    
\newcount\countno                  
\newcount\refno                    
\newcount\eqlett     \eqlett=97    
\newif\ifletter 
\newif\ifwide 
\newif\ifnotfull 
\newif\ifaligned 
\newif\ifnumbysec   
\newif\ifappendix 
\newif\ifnumapp 
\newif\ifssf 
\newif\ifppt 
\newdimen\t@bwidth 
\newdimen\c@pwidth 
\newdimen\digitwidth                    
\newdimen\argwidth                      
\newdimen\secindent    \secindent=5pc   
\newdimen\textind    \textind=16pt      
\newdimen\tempval                       
\newskip\beforesecskip 
\def\beforesecspace{\vskip\beforesecskip\relax} 
\newskip\beforesubskip 
\def\beforesubspace{\vskip\beforesubskip\relax} 
\newskip\beforesubsubskip 
\def\beforesubsubspace{\vskip\beforesubsubskip\relax} 
\newskip\secskip 
\def\secspace{\vskip\secskip\relax} 
\newskip\subskip 
\def\subspace{\vskip\subskip\relax} 
\newskip\insertskip 
\def\insertspace{\vskip\insertskip\relax} 
\def\sp@ce{\ifx\next*\let\next=\@ssf 
               \else\let\next=\@nossf\fi\next} 
\def\@ssf#1{\nobreak\secspace\global\ssftrue\nobreak} 
\def\@nossf{\nobreak\secspace\nobreak\noindent\ignorespaces} 
\def\subsp@ce{\ifx\next*\let\next=\@sssf 
               \else\let\next=\@nosssf\fi\next} 
\def\@sssf#1{\nobreak\subspace\global\ssftrue\nobreak} 
\def\@nosssf{\nobreak\subspace\nobreak\noindent\ignorespaces} 
\beforesecskip=24pt plus12pt minus8pt 
\beforesubskip=12pt plus6pt minus4pt 
\beforesubsubskip=12pt plus6pt minus4pt 
\secskip=12pt plus 2pt minus 2pt 
\subskip=6pt plus3pt minus2pt 
\insertskip=18pt plus6pt minus6pt%
\fontdimen16\tensy=2.7pt 
\fontdimen17\tensy=2.7pt 
%
%
\def\eqlabel{(\ifappendix\applett 
               \ifnumbysec\ifnum\secno>0 \the\secno\fi.\fi 
               \else\ifnumbysec\the\secno.\fi\fi\the\countno)} 
\def\seclabel{\ifappendix\ifnumapp\else\applett\fi 
    \ifnum\secno>0 \the\secno 
    \ifnumbysec\ifnum\subno>0.\the\subno\fi\fi\fi 
    \else\the\secno\fi\ifnum\subno>0.\the\subno 
         \ifnum\subsubno>0.\the\subsubno\fi\fi} 
\def\tablabel{\ifappendix\applett\fi\the\tabno} 
\def\figlabel{\ifappendix\applett\fi\the\figno} 
\def\gac{\global\advance\countno by 1} 
%
%
 
\def\vfootnote#1{\insert\footins\bgroup 
\interlinepenalty=\interfootnotelinepenalty 
\splittopskip=\ht\strutbox 
\splitmaxdepth=\dp\strutbox \floatingpenalty=20000 
\leftskip=0pt \rightskip=0pt \spaceskip=0pt \xspaceskip=0pt%
\noindent\smallfonts\rm #1\ \ignorespaces\footstrut\futurelet\next\fo@t} 
%
%
\def\endinsert{\egroup 
    \if@mid \dimen@=\ht0 \advance\dimen@ by\dp0 
       \advance\dimen@ by12\p@ \advance\dimen@ by\pagetotal 
       \ifdim\dimen@>\pagegoal \@midfalse\p@gefalse\fi\fi 
    \if@mid \insertspace \box0 \par \ifdim\lastskip<\insertskip 
    \removelastskip \penalty-200 \insertspace \fi 
    \else\insert\topins{\penalty100 
       \splittopskip=0pt \splitmaxdepth=\maxdimen  
       \floatingpenalty=0 
       \ifp@ge \dimen@=\dp0 
       \vbox to\vsize{\unvbox0 \kern-\dimen@}%
       \else\box0\nobreak\insertspace\fi}\fi\endgroup}    
%
%
%
\def\ind{\hbox to \secindent{\hfill}} 
%
%

%
%
 
%
%
\def\indeqn#1{\alignedfalse\displ@y\halign{\hbox to \displaywidth 
    {$\ind\@lign\displaystyle##\hfil$}\crcr #1\crcr}} 
%
%
\def\indalign#1{\alignedtrue\displ@y \tabskip=0pt  
  \halign to\displaywidth{\ind$\@lign\displaystyle{##}$\tabskip=0pt 
    &$\@lign\displaystyle{{}##}$\hfill\tabskip=\centering 
    &\llap{$\@lign\hbox{\rm##}$}\tabskip=0pt\crcr 
    #1\crcr}} 
\def\indenteddisplay#1$${\indispl@y{#1 }} 
\def\indispl@y#1{\disptest#1\eqalignno\eqalignno\disptest} 
\def\disptest#1\eqalignno#2\eqalignno#3\disptest{%
    \ifx#3\eqalignno 
    \indalign#2%
    \else\indeqn{#1}\fi$$} 
%
%
 
%
%
 
%
%
 
%
%
 
%
%

\def\ns{\noalign{\vskip-3pt}}

%
 
%
%
\def\bhbar{\rlap{\kern1pt\raise.4ex\hbox{\bf\char'40}}\bi{h}} 

\def\d{{\rm d}}

\def\frac#1#2{{#1\over#2}} 
\ifams 
\def\lap{\lesssim} 
\def\gap{\gtrsim}

\let\leq=\leqslant

\else

\def\gap{\;\lower3pt\hbox{$\buildrel > \over \sim$}\;}%
\def\lap{\;\lower3pt\hbox{$\buildrel < \over \sim$}\;}\fi 
 
\chardef\ii="10 
\def\tqs{\hbox to 25pt{\hfil}}

\def\Bbbone{1\kern-.22em {\rm l}} 
%
%
\def\rp{\raise8pt\hbox{$\scriptstyle\prime$}} 
%
%
%
%

%
%
\def\[#1\]{\setbox0=\hbox{$\dsty#1$}\argwidth=\wd0 
    \setbox0=\hbox{$\left[\box0\right]$}\advance\argwidth by -\wd0 
    \left[\kern.3\argwidth\box0\kern.3\argwidth\right]} 
%
%
\def\lsb#1\rsb{\setbox0=\hbox{$#1$}\argwidth=\wd0 
    \setbox0=\hbox{$\left[\box0\right]$}\advance\argwidth by -\wd0 
    \left[\kern.3\argwidth\box0\kern.3\argwidth\right]} 
%
 
%
%
 
%
\def\pt(#1){({\it #1\/})} 
\let\dsty=\displaystyle

%
%
\def\reactions#1{\vskip 12pt plus2pt minus2pt%
\vbox{\hbox{\kern\secindent\vrule\kern12pt%
\vbox{\kern0.5pt\vbox{\hsize=24pc\parindent=0pt\smallfonts\rm NUCLEAR  
REACTIONS\strut\quad #1\strut}\kern0.5pt}\kern12pt\vrule}}} 
%
%
\def\slashchar#1{\setbox0=\hbox{$#1$}\dimen0=\wd0%
\setbox1=\hbox{/}\dimen1=\wd1%
\ifdim\dimen0>\dimen1%
\rlap{\hbox to \dimen0{\hfil/\hfil}}#1\else                                         
\rlap{\hbox to \dimen1{\hfil$#1$\hfil}}/\fi} 
%
%
\def\textindent#1{\noindent\hbox to \parindent{#1\hss}\ignorespaces} 
%
%
\def\opencirc{\raise1pt\hbox{$\scriptstyle{\bigcirc}$}} 
 
\ifams 
\def\opensqr{\hbox{$\square$}} 
 
\def\opentridown{\hbox{$\triangledown$}}

\else 
\def\opensqr{\vbox{\hrule height.4pt\hbox{\vrule width.4pt height3.5pt 
    \kern3.5pt\vrule width.4pt}\hrule height.4pt}} 
 
\def\opentridown{\raise1pt\hbox{$\scriptstyle\bigtriangledown$}}

\fi

%
%
\def\m@th{\mathsurround=0pt} 
%
%
\def\cases#1{%
\left\{\,\vcenter{\normalbaselines\openup1\jot\m@th%
     \ialign{$\displaystyle##\hfil$&\rm\tqs##\hfil\crcr#1\crcr}}\right.}%
%
%
\def\oldcases#1{\left\{\,\vcenter{\normalbaselines\m@th 
    \ialign{$##\hfil$&\rm\quad##\hfil\crcr#1\crcr}}\right.} 
%
%
\def\numcases#1{\left\{\,\vcenter{\baselineskip=15pt\m@th%
     \ialign{$\displaystyle##\hfil$&\rm\tqs##\hfil 
     \crcr#1\crcr}}\right.\hfill 
     \vcenter{\baselineskip=15pt\m@th%
     \ialign{\rlap{$\phantom{\displaystyle##\hfil}$}\tabskip=0pt&\en 
     \rlap{\phantom{##\hfil}}\crcr#1\crcr}}} 
\def\ptnumcases#1{\left\{\,\vcenter{\baselineskip=15pt\m@th%
     \ialign{$\displaystyle##\hfil$&\rm\tqs##\hfil 
     \crcr#1\crcr}}\right.\hfill 
     \vcenter{\baselineskip=15pt\m@th%
     \ialign{\rlap{$\phantom{\displaystyle##\hfil}$}\tabskip=0pt&\enpt 
     \rlap{\phantom{##\hfil}}\crcr#1\crcr}}\global\eqlett=97 
     \global\advance\countno by 1} 
%
%
\def\eq(#1){\ifaligned\@mp(#1)\else\hfill\llap{{\rm (#1)}}\fi} 
\def\ceq(#1){\ns\ns\ifaligned\@mp\fi\eq(#1)\cr\ns\ns} 
\def\eqpt(#1#2){\ifaligned\@mp(#1{\it #2\/}) 
                    \else\hfill\llap{{\rm (#1{\it #2\/})}}\fi} 
\let\eqno=\eq 
%
%
\countno=1 
 
\def\aleq{&\rm(\ifappendix\applett 
               \ifnumbysec\ifnum\secno>0 \the\secno\fi.\fi 
               \else\ifnumbysec\the\secno.\fi\fi\the\countno} 
\def\noaleq{\hfill\llap\bgroup\rm(\ifappendix\applett 
               \ifnumbysec\ifnum\secno>0 \the\secno\fi.\fi 
               \else\ifnumbysec\the\secno.\fi\fi\the\countno} 
\def\@mp{&} 
\def\en{\ifaligned\aleq)\else\noaleq)\egroup\fi\gac} 
\def\cen{\ns\ns\ifaligned\@mp\fi\en\cr\ns\ns} 
\def\enpt{\ifaligned\aleq{\it\char\the\eqlett})\else 
    \noaleq{\it\char\the\eqlett})\egroup\fi 
    \global\advance\eqlett by 1} 
\def\endpt{\ifaligned\aleq{\it\char\the\eqlett})\else 
    \noaleq{\it\char\the\eqlett})\egroup\fi 
    \global\eqlett=97\gac} 
%
%


 

%
%

\def\PRL{{\it Phys. Rev. Lett.}}

\def\ZP{{\it Z. Phys.}} 
\headline={\ifodd\pageno{\ifnum\pageno=\firstpage\hfill 
   \else\rrhead\fi}\else\lrhead\fi} 
\def\rrhead{\textfonts\hskip\secindent\it 
    \shorttitle\hfill\rm\folio} 
\def\lrhead{\textfonts\hbox to\secindent{\rm\folio\hss}%
    \it\aunames\hss} 
\footline={\ifnum\pageno=\firstpage \hfill\textfonts\rm\folio\fi} 
\def\@rticle#1#2{\vglue.5pc 
    {\parindent=\secindent \bf #1\par} 
     \vskip2.5pc 
    {\exhyphenpenalty=10000\hyphenpenalty=10000 
     \baselineskip=18pt\raggedright\noindent 
     \headfonts\bf#2\par}\futurelet\next\sh@rttitle}%
\def\title#1{\gdef\shorttitle{#1} 
    \vglue4pc{\exhyphenpenalty=10000\hyphenpenalty=10000  
    \baselineskip=18pt  
    \raggedright\parindent=0pt 
    \headfonts\bf#1\par}\futurelet\next\sh@rttitle}  

\def\article#1#2{\gdef\shorttitle{#2}\@rticle{#1}{#2}}  
\def\review#1{\gdef\shorttitle{#1}%
    \@rticle{REVIEW \ifpbm\else ARTICLE\fi}{#1}} 
\def\topical#1{\gdef\shorttitle{#1}%
    \@rticle{TOPICAL REVIEW}{#1}} 
\def\comment#1{\gdef\shorttitle{#1}%
    \@rticle{COMMENT}{#1}} 
\def\note#1{\gdef\shorttitle{#1}%
    \@rticle{NOTE}{#1}} 
\def\prelim#1{\gdef\shorttitle{#1}%
    \@rticle{PRELIMINARY COMMUNICATION}{#1}} 
\def\letter#1{\gdef\shorttitle{Letter to the Editor}%
     \gdef\aunames{Letter to the Editor} 
     \global\lettertrue\ifnum\jnl=7\global\letterfalse\fi 
     \@rticle{LETTER TO THE EDITOR}{#1}} 
\def\sh@rttitle{\ifx\next[\let\next=\sh@rt 
                \else\let\next=\f@ll\fi\next} 
\def\sh@rt[#1]{\gdef\shorttitle{#1}} 
\def\f@ll{} 
\def\author#1{\ifletter\else\gdef\aunames{#1}\fi\vskip1.5pc 
    {\parindent=\secindent   
     \hang\textfonts   
     \ifppt\bf\else\rm\fi#1\par}   
     \ifppt\bigskip\else\smallskip\fi 
     \futurelet\next\@unames} 
\def\@unames{\ifx\next[\let\next=\short@uthor 
                 \else\let\next=\@uthor\fi\next} 
\def\short@uthor[#1]{\gdef\aunames{#1}} 
\def\@uthor{} 
\def\address#1{{\parindent=\secindent 
    \exhyphenpenalty=10000\hyphenpenalty=10000 
\ifppt\textfonts\else\smallfonts\fi\hang\raggedright\rm#1\par}%
    \ifppt\bigskip\fi} 
\def\jl#1{\global\jnl=#1} 
\jl{0}%
\def\journal{\ifnum\jnl=1 J. Phys.\ A: Math.\ Gen.\  
        \else\ifnum\jnl=2 J. Phys.\ B: At.\ Mol.\ Opt.\ Phys.\  
        \else\ifnum\jnl=3 J. Phys.:\ Condens. Matter\  
        \else\ifnum\jnl=4 J. Phys.\ G: Nucl.\ Part.\ Phys.\  
        \else\ifnum\jnl=5 Inverse Problems\  
        \else\ifnum\jnl=6 Class. Quantum Grav.\  
        \else\ifnum\jnl=7 Network\  
        \else\ifnum\jnl=8 Nonlinearity\ 
        \else\ifnum\jnl=9 Quantum Opt.\ 
        \else\ifnum\jnl=10 Waves in Random Media\ 
        \else\ifnum\jnl=11 Pure Appl. Opt.\  
        \else\ifnum\jnl=12 Phys. Med. Biol.\ 
        \else\ifnum\jnl=13 Modelling Simulation Mater.\ Sci.\ Eng.\  
        \else\ifnum\jnl=14 Plasma Phys. Control. Fusion\  
        \else\ifnum\jnl=15 Physiol. Meas.\  
        \else\ifnum\jnl=16 Sov.\ Lightwave Commun.\ 
        \else\ifnum\jnl=17 J. Phys.\ D: Appl.\ Phys.\ 
        \else\ifnum\jnl=18 Supercond.\ Sci.\ Technol.\ 
        \else\ifnum\jnl=19 Semicond.\ Sci.\ Technol.\ 
        \else\ifnum\jnl=20 Nanotechnology\ 
        \else\ifnum\jnl=21 Meas.\ Sci.\ Technol.\  
        \else\ifnum\jnl=22 Plasma Sources Sci.\ Technol.\  
        \else\ifnum\jnl=23 Smart Mater.\ Struct.\  
        \else\ifnum\jnl=24 J.\ Micromech.\ Microeng.\ 
   \else Institute of Physics Publishing\  
   \fi\fi\fi\fi\fi\fi\fi\fi\fi\fi\fi\fi\fi\fi\fi 
   \fi\fi\fi\fi\fi\fi\fi\fi\fi} 
\let\abs=\beginabstract 

\let\endabs=\endabstract 
\def\today{\number\day\ \ifcase\month\or 
     January\or February\or March\or April\or May\or June\or 
     July\or August\or September\or October\or November\or 
     December\fi\space \number\year} 
\def\date{\ifppt\noindent\textfonts\rm  
     Date: \today\par\goodbreak\bigskip\fi} 
%
%
 
%
 
%
%
\def\section#1{\ifppt\ifnum\secno=0\eject\fi\fi 
    \subno=0\subsubno=0\global\advance\secno by 1 
    \gdef\labeltype{\seclabel}\ifnumbysec\countno=1\fi 
    \goodbreak\beforesecspace\nobreak 
    \noindent{\bf \the\secno. #1}\par\futurelet\next\sp@ce} 
\def\subsection#1{\subsubno=0\global\advance\subno by 1 
     \gdef\labeltype{\seclabel}%
     \ifssf\else\goodbreak\beforesubspace\fi 
     \global\ssffalse\nobreak 
     \noindent{\it \the\secno.\the\subno. #1\par}%
     \futurelet\next\subsp@ce} 
\def\subsubsection#1{\global\advance\subsubno by 1 
     \gdef\labeltype{\seclabel}%
     \ifssf\else\goodbreak\beforesubsubspace\fi 
     \global\ssffalse\nobreak 
     \noindent{\it \the\secno.\the\subno.\the\subsubno. #1}\null.  
     \ignorespaces} 
%
 
%
%
\def\numappendix#1{\ifappendix\ifnumbysec\countno=1\fi\else 
    \countno=1\figno=0\tabno=0\fi 
    \subno=0\global\advance\appno by 1 
    \secno=\appno\gdef\applett{A}\gdef\labeltype{\seclabel}%
    \global\appendixtrue\global\numapptrue 
    \goodbreak\beforesecspace\nobreak 
    \noindent{\bf Appendix \the\appno. #1\par}%
    \futurelet\next\sp@ce} 
\def\numsubappendix#1{\global\advance\subno by 1\subsubno=0 
    \gdef\labeltype{\seclabel}%
    \ifssf\else\goodbreak\beforesubspace\fi 
    \global\ssffalse\nobreak 
    \noindent{\it A\the\appno.\the\subno. #1\par}%
    \futurelet\next\subsp@ce} 
\def\@ppendix#1#2#3{\countno=1\subno=0\subsubno=0\secno=0\figno=0\tabno=0 
    \gdef\applett{#1}\gdef\labeltype{\seclabel}\global\appendixtrue 
    \goodbreak\beforesecspace\nobreak 
    \noindent{\bf Appendix#2#3\par}\futurelet\next\sp@ce} 
\def\Appendix#1{\@ppendix{A}{. }{#1}} 
\def\appendix#1#2{\@ppendix{#1}{ #1. }{#2}} 
\def\App#1{\@ppendix{A}{ }{#1}} 
\def\app{\@ppendix{A}{}{}} 
\def\subappendix#1#2{\global\advance\subno by 1\subsubno=0 
    \gdef\labeltype{\seclabel}%
    \ifssf\else\goodbreak\beforesubspace\fi 
    \global\ssffalse\nobreak 
    \noindent{\it #1\the\subno. #2\par}%
    \nobreak\subspace\noindent\ignorespaces} 
%
%
\def\@ck#1{\ifletter\bigskip\noindent\ignorespaces\else 
    \goodbreak\beforesecspace\nobreak 
    \noindent{\bf Acknowledgment#1\par}%
    \nobreak\secspace\noindent\ignorespaces\fi} 
\def\ack{\@ck{s}} 
\def\ackn{\@ck{}} 
\def\n@ip#1{\goodbreak\beforesecspace\nobreak 
    \noindent\smallfonts{\it #1}. \rm\ignorespaces} 
\def\naip{\n@ip{Note added in proof}} 
\def\na{\n@ip{Note added}} 
 
%
%
 
%
 
%
%
 
%
 
%
 
\def\tablecont{\topinsert\global\advance\tabno by -1 
    \tablecaption{(continued)}} 
\def\tablecaption#1{\gdef\labeltype{\tablabel}\global\widefalse 
    \leftskip=\secindent\parindent=0pt 
    \global\advance\tabno by 1 
    \smallfonts{\bf Table \ifappendix\applett\fi\the\tabno.} \rm #1\par 
    \smallskip\futurelet\next\t@b} 
\def\t@b{\ifx\next*\let\next=\widet@b 
             \else\ifx\next[\let\next=\fullwidet@b 
                      \else\let\next=\narrowt@b\fi\fi 
             \next} 
\def\widet@b#1{\global\widetrue\global\notfulltrue 
    \t@bwidth=\hsize\advance\t@bwidth by -\secindent}  
\def\fullwidet@b[#1]{\global\widetrue\global\notfullfalse 
    \leftskip=0pt\t@bwidth=\hsize}                   
\def\narrowt@b{\global\notfulltrue} 
\def\align{\catcode`?=13\ifnotfull\moveright\secindent\fi 
    \vbox\bgroup\halign\ifwide to \t@bwidth\fi 
    \bgroup\strut\tabskip=1.2pc plus1pc minus.5pc} 
\def\endalign{\egroup\egroup\catcode`?=12} 
 
%
%

%
%

%
 
%
%

%
 
\catcode`?=13 
\def\lineup{\setbox0=\hbox{\smallfonts\rm 0}%
    \digitwidth=\wd0%
    \def?{\kern\digitwidth}%
    \def\\{\hbox{$\phantom{-}$}}%
    \def\-{\llap{$-$}}} 
\catcode`?=12 
%
%
\def\sidetable#1#2{\hbox{\ifppt\hsize=18pc\t@bwidth=18pc 
                          \else\hsize=15pc\t@bwidth=15pc\fi 
    \parindent=0pt\vtop{\null #1\par}%
    \ifppt\hskip1.2pc\else\hskip1pc\fi 
    \vtop{\null #2\par}}}  
\def\lstable#1#2{\everypar{}\tempval=\hsize\hsize=\vsize 
    \vsize=\tempval\hoffset=-3pc 
    \global\tabno=#1\gdef\labeltype{\tablabel}%
    \noindent\smallfonts{\bf Table \ifappendix\applett\fi 
    \the\tabno.} \rm #2\par 
    \smallskip\futurelet\next\t@b} 
\def\inctabno{\global\advance\tabno by 1} 
%
%
 
%
 
%
\def\figure#1{\figc@ption{#1}\bigskip} 
\def\figc@ption#1{\global\advance\figno by 1\gdef\labeltype{\figlabel}%
   {\parindent=\secindent\smallfonts\hang 
    {\bf Figure \ifappendix\applett\fi\the\figno.} \rm #1\par}} 
%
%
\def\refHEAD{\goodbreak\beforesecspace 
     \noindent\textfonts{\bf References}\par 
     \let\ref=\rf 
     \nobreak\smallfonts\rm} 
\def\references{\refHEAD\parindent=0pt 
     \everypar{\hangindent=18pt\hangafter=1 
     \frenchspacing\rm}%
     \secspace} 
\def\rf#1{\par\noindent\hbox to 21pt{\hss #1\quad}\ignorespaces} 
%
 
%
 
%
%
\def\numrefjl#1#2#3#4#5{\par\rf{#1}#2 {\it #3 \bf #4} #5\par} 
%
%
 
%
%

%
%

%
\catcode`\@=12 
%
%
 
%
%
\def\jnlstyle{\pptfalse\headsize{14}{18}%
\textsize{10}{12}%
\smallsize{8}{10} 
\textind=16pt} 
%
%
 
%
%
 
%
\parindent=\textind 
%
\input epsf.tex
\def\ie{{\it i.e.} }
\def\bra#1{\langle#1\vert}
\def\ket#1{\vert#1\rangle}
\def\eg{{\it e.g.} }
\def\mod{{\rm mod}}
\headline={\ifodd\pageno{\ifnum\pageno=\firstpage\titlehead
   \else\rrhead\fi}\else\lrhead\fi} 
\def\lpsn#1#2{LPSN-#1-LT#2}
\footline={\ifnum\pageno=\firstpage{\smallfonts cond--mat/9306013}
\hfil\textfonts\rm\folio\fi}   
\def\titlehead{\smallfonts \ZP{\ }B\ {\bf 92} (1993) 307 
\hfil\lpsn{93}{2}} 

\jnlstyle
\jl{1}
\overfullrule=0pt

\title{Surface magnetization of aperiodic Ising quantum chains}
\author{L Turban and B Berche}
 
\address{Laboratoire de Physique du Solide, URA CNRS $n^o155$, 
Universit\'e de Nancy I, BP239, F-54506 
Vand\oe uvre l\`es Nancy Cedex, France}
 
\abs
We study the surface magnetization of aperiodic Ising quantum chains.
Using fermion techniques, exact results are obtained
in the critical region for quasiperiodic sequences generated
through an irrational number as well as for the automatic binary Thue-Morse 
sequence and its generalizations modulo $p$. The surface 
magnetization exponent keeps its Ising value, $\beta_s=1/2$, for 
all the sequences studied. The critical amplitude of the surface 
magnetization depends on the strength of the modulation
and also on the starting point of the chain along the aperiodic
sequence.   
\endabs
\section{Introduction}
Since the discovery of quasicrystals [1] there has been a growing
interest in the structural and physical properties of quasiperiodic
systems (see \eg [2-5]). Due to their situation, which is 
intermediate between periodic and random ones, these systems are
particularly interesting from the point of view of their 
critical behaviour. 

On the Penrose lattice, an approximate renormalization group
study of the ferromagnetic Ising model [6] suggested a 
logarithmic singularity in the specific heat, \ie ordinary
two-dimensional Ising behaviour. In the same way, random and 
self-avoiding walks, studied via Monte-Carlo techniques, were
shown to keep their usual asymptotic properties [7].

Exact results have been obtained, either for aperiodic quantum 
spin chains or for the corresponding layered classical 
counterparts ([8-12] and references therein). A condition for 
quasiperiodicity was established [13] in the case of aperiodic chains 
generated through an inflation rule. The one-dimensional lattice
is quasiperiodic, \ie it displays a Fourier spectrum with Dirac
delta peaks for a set of $Q$-values which are not integral 
multiples of any single wave-vector, when the characteristic
polynomial of the substitution has only one root with an
absolute value greater than one.
 
Mainly bulk critical properties of aperiodic Ising systems 
have been 
worked out in the last years. Tracy [9] found that, although for
quasiperiodically layered Ising systems the specific heat displays 
a logarithmic singularity, this is no longer true for some aperiodic 
sequences which do not satisfy the criterion for quasiperiodicity.
The low energy finite-size behaviour of the spectrum of 
quasiperiodic Ising quantum chains was found to be in agreement
with the prediction of conformal theory by Igl\'oi [10]. The bulk
critical exponents deduced from the gaps are those of the 
periodic system. The structure of the finite-size Ising spectrum 
has been carefully studied for the Fibonacci chain in [12]. Besides
the conformally invariant scaling of the levels at low energy, a
multifractal scaling is observed for higher ones.

In the present work we examine the surface critical behaviour of
aperiodic Ising quantum chains. The surface magnetization is
calculated exactly in the critical region making use of fermion 
techniques [14,15]. The outline of the paper is as follows. In
section $2$ we recall how the surface magnetization of the
inhomogeneous Ising quantum chain can be obtained. The case of
a quasiperiodic chain generated by an irrational number is considered 
in section $3$. This is followed by a study of the aperiodic binary 
Thue-Morse sequence in section $4$ and of its generalizations
modulo $p$ in section $5$. The results are discussed in
section $6$.
\section{Surface magnetization of the inhomogeneous Ising quantum
chain}
Let us consider a semi-infinite inhomogeneous Ising quantum chain 
with modulated interaction strength $\lambda_j$ and a constant 
transverse field $h\!=\!1$. The Hamiltonian can be written as
$$
{\cal H}=-{1\over 2}\sum_{j=1}^\infty\left[\sigma_j^z+\lambda_j
\sigma_j^x\sigma_{j+1}^x\right]\eqno(2.1)
$$
where the $\sigma$'s are Pauli spin operators. This quantum chain
also corresponds, in the extreme anisotropic limit, to a
semi-infinite layered two-dimensional classical Ising system
with modulated interactions on the bonds perpendicular to the
surface.

Using the Jordan-Wigner transformation [16] the Hamiltonian is
rewritten as a quadratic form in fermion operators $c_j^+$, 
$c_j$, which may be diagonalized through the canonical 
transformation [14,17] 
$$
c_j={1\over 2}\sum_\alpha\left\{\left[\Phi_\alpha(j)+\Psi_\alpha
(j)\right]\eta_\alpha+\left[\Phi_\alpha(j)-\Psi_\alpha(j)\right]
\eta_\alpha^+\right\}\eqno(2.2)
$$
where $\eta_\alpha$ and $\eta_\alpha^+$ are new fermion 
operators. ${\bf \Phi}_\alpha$ and ${\bf \Psi}_\alpha$ are the
real normalized eigenvectors of two matrices leading to the 
excitation energies $\epsilon_\alpha$ of the diagonal Hamiltonian
${\cal H}=\sum_\alpha\epsilon_\alpha(\eta_\alpha^+
\eta_\alpha-1/2)$. They are related through
$$
(\bss{A}+\bss{B}){\bf\Phi}_\alpha=\epsilon_\alpha{\bf\Psi}_\alpha
\qquad(\bss{A}+\bss{B})_{ij}=-\delta_{ij}-\lambda_i\delta_{i+1,j}
\eqno(2.3)
$$
Below the critical point, in the ordered phase, the ground-state 
is two-fold degenerate and the lowest excitation $\epsilon_0$, 
which corresponds to an eigenvector localized near to the surface, 
vanishes exponentially with the size of the system. This
provides the condition for criticality [18]
$$
\lim_{N\rightarrow\infty}\prod_{j=1}^N(\lambda_j)_c=1\eqno(2.4)
$$
The surface magnetization can be obtained using a simple 
method due to Peschel [15]. The asymptotic form of the
surface two-point correlation function gives the square of the
surface magnetization $m_s$ which can be expressed as the matrix 
element $\bra{0}\sigma_1^x\ket{\sigma}$ where 
$\ket{0}$ is the 
vacuum of ${\cal H}$ and $\ket{\sigma}\!=\!\eta_0^+\ket{0}$ its first 
excited state. Rewriting $\sigma_1^x$ in terms of diagonal 
fermions through (2.2), one obtains $m_s\!=\!\Phi_0(1)$. Using (2.3) 
with $\alpha\!=\!0$ in the ordered phase where the lowest excitation 
vanishes, a recursion equation for the successive components of 
${\bf\Phi}_0$ is obtained leading to
$$
\Phi_0(j)=(-1)^{j-1}\Phi_0(1)\prod_{i=1}^{j-1}\lambda_i^{-1}\qquad
j>1\eqno(2.5)
$$
The surface magnetization then follows from the normalization of
the eigenvector and reads
$$
m_s=\left(1+\sum_{j=1}^\infty\prod_{i=1}^j\lambda_i^{-2}
\right)^{-1/2}\eqno(2.6)
$$
\section{Quasiperiodic modulation generated by an irrational number}
Let the position $x_j$ of the $j$-th spin along the chain be
defined as [10, 13]
$$
x_j=jb+\left[{j+1\over\omega}\right](a-b)\qquad(a>b)\eqno(3.1)
$$
where $[x]$ is the integer part of $x$. When $\omega$ is 
irrational, one obtains a succession of short and long bonds
which are distributed according to a quasiperiodic sequence.
With a modulation amplitude $r$\footnote{\dag}{In the following
we take $r>0$ without loss of generality.} such that $a$-bonds 
have a strength $\lambda$ and $b$-bonds a strength $\lambda r$, 
the number of modified bonds from the surface to site $j$ is
given by
$$
n_j=j-\left[{j+1\over \omega}\right]\eqno(3.2)
$$
corresponding to an asymptotic density $\rho_\infty\!=\!1\!-
\!1/\omega$.
The critical coupling $\lambda_c$ follows from (2.4) with
$$
\lim_{N\rightarrow\infty}\prod_{j=1}^N(\lambda_j)_c=\lim_{N
\rightarrow\infty}\lambda_c^Nr^{N\rho_\infty}=1\eqno(3.3)
$$
so that
$$
\lambda_c=r^{1/\omega -1}\eqno(3.4)
$$
The surface magnetization (2.6) involves the sum of products
$$
\sum_{j=1}^\infty\prod_{i=1}^j\lambda_i^{-2}=\sum_{j=1}^\infty
\lambda^{-2j}r^{-2n_j}=r^{2/\omega}\sum_{j=1}^\infty\left(
{\lambda_c\over\lambda}\right)^{2j}r^{-2\left\{{j+1\over\omega}
\right\}}\eqno(3.5)
$$
In the last term $n_j$, given by (3.2), was rewritten as
$$
n_j=j-{j+1\over\omega}+\left\{{j+1\over\omega}\right\}\eqno(3.6)
$$
where $\{x\}$ is the fractional part of $x$. The eigenvector 
${\bf\Phi}_0$, which is localized near to the surface in the 
ordered
phase, extends into the bulk in the critical regime ($\lambda
\!\rightarrow\!\lambda_c$). Then the last factor in (3.5) may be 
replaced by its averaged value [11]. Since $\omega$ is irrational,
$\left\{(j\!+\!1)/\omega\right\}$ is uniformly distributed over
$[0,1]$ and the average is given by
$$
\left<r^{-2\left\{{j+1\over\omega}\right\}}\right>=\int_0^1
r^{-2x}\d x={r^2-1\over r^2\ln r^2}\eqno(3.7)
$$
Summing the geometrical series, one obtains the exact leading 
contribution in the scaling region
$$
m_s^{(1)}={1\over\lambda_c}\left[{\ln r^2\over r^2-1}\right]^{1/2}
\left[1-\left({\lambda_c\over\lambda}\right)^2\right]^{1/2}
\eqno(3.8)
$$
The surface exponent $\beta_s\!=\!1/2$ is the same as for an
homogeneous Ising system and the critical amplitude keeps the
same form for any sequence generated by an irrational. It 
only depends on the modulation amplitude and the critical 
coupling.

{\par\begingroup\parindent=0pt\medskip
\epsfxsize=9truecm
\topinsert
\centerline{\epsfbox{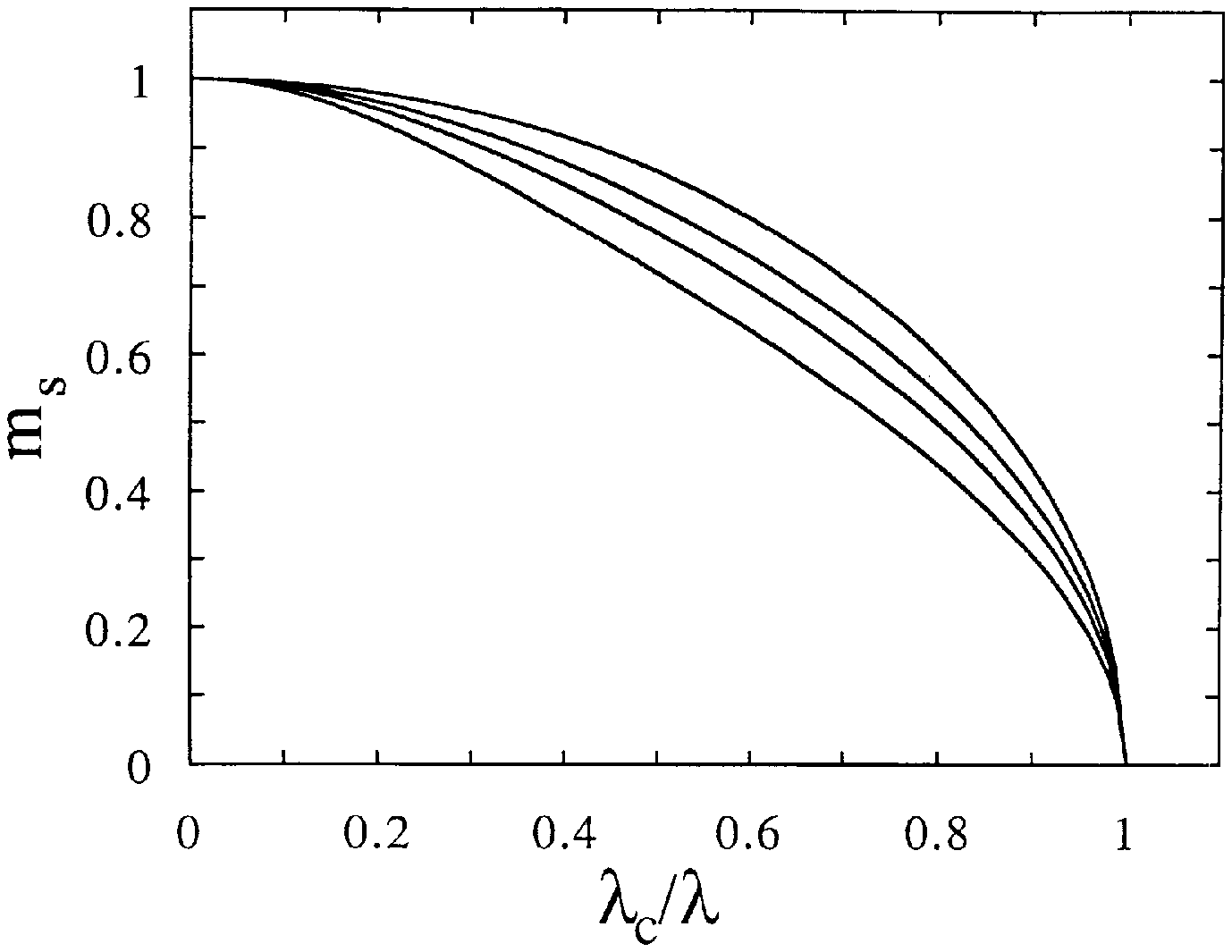}}
\smallskip
\figure{Surface magnetization $m_s$ of the Fibonacci Ising 
quantum chain with a modulation amplitude $r=1,2,3,5$ from top
to bottom.}
\endinsert 
\endgroup
\par}

{\par\begingroup\parindent=0pt\medskip
\epsfxsize=9truecm
\topinsert
\centerline{\epsfbox{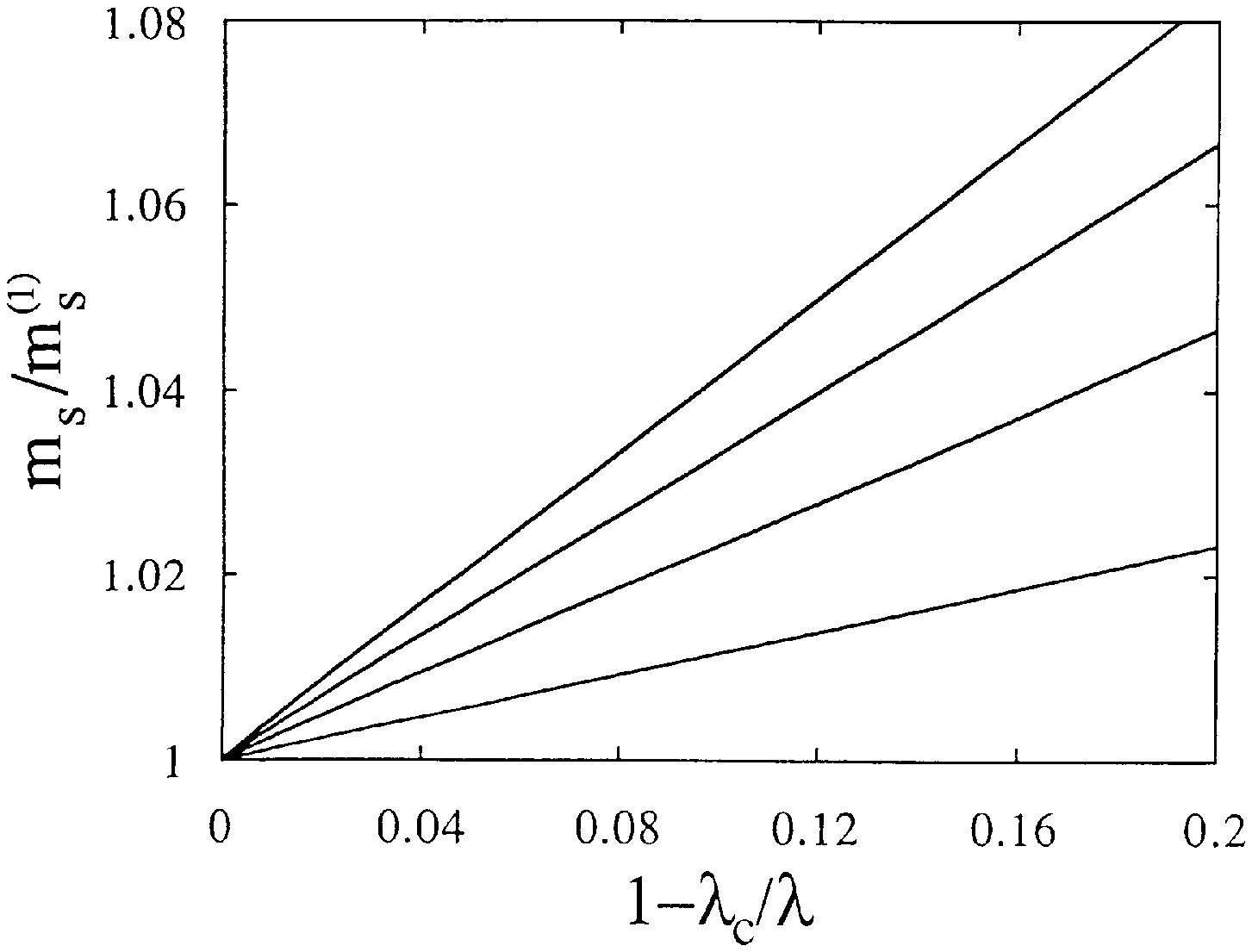}}
\smallskip
\figure{Ratio of the Fibonacci surface magnetization $m_s$ to
its leading critical term $m_s^{(1)}\sim\epsilon^{1/2}$ for 
$r=2,3,4,5$ from bottom to top. The linear behaviour corresponds
to a correction to scaling of order $\epsilon^{3/2}$.}
\endinsert 
\endgroup
\par}

When $\omega$ is the golden mean $(\sqrt{5}\!+\!1)/2$, the modulation
is distributed according to the Fibonacci sequence. Such a 
sequence may be obtained in various ways. It can be represented 
by an infinite word $W_\infty\!=\!01001010\dots$ where $0$ stands
for an $a$-bond and $1$ for a $b$-bond. This word is the limit 
of an infinite sequence of finite words $W_n$ obtained through 
the recursion relation 
$$
W_{n+1}=W_nW_{n-1}\qquad W_0=1\qquad W_1=0\eqno(3.9)
$$
to which is associated the characteristic polynomial 
$x^2-x-1$ [9]. The length of succesive words is given by  
Fibonacci numbers.

The Fibonacci sequence also follows from the inflation rule [9]
$$
0\rightarrow 01\qquad 1\rightarrow 0\eqno(3.10)
$$
giving
$$
\eqalign{&0\cr&01\cr&010\cr&01001\cr&01001010\cr&\dots}\eqno(3.11)
$$

The surface magnetization of the Fibonacci quasilattice 
as given by (2.6) is plotted in figure $1$ for different values of
the modulation amplitude $r$. The ratio $m_s/m_s^{(1)}$ shown in
figure $2$ as a function of $\epsilon=1-\lambda_c/\lambda$ indicates
that the corrections to scaling are of order $\epsilon^{3/2}$ in 
this case.

\section{The binary Thue-Morse sequence}
Let us now consider the binary Thue-Morse sequence $f_j^{(2)}=0$
or $1$ $(j\!=\!1,2,3\dots)$
which {\it is not quasiperiodic} [19] and sometimes called 
automatic since it can be generated through a $2$-automaton. It may 
be deduced from the binary representation of non-negative integers
[20]
$$
0\quad 1\quad 10\quad 11\quad 100\quad 101\quad 110\quad 111
\quad\dots\eqno(4.1)
$$
by counting the sum of their digits modulo $2$
$$
01101001\dots\eqno(4.2)
$$
or through the recursion
$$
W_{n+1}=W_n\left(W_{n-1}^{-1}W_n\right)W_{n-1}\eqno(4.3)
$$
It also results from the substitution
$$
0\rightarrow 01\qquad 1\rightarrow 10\eqno(4.4)
$$
leading to
$$
\eqalign{&0\cr&01\cr&0110\cr&01101001\cr&\underline{0}1
\underline{1}0\underline{1}0\underline{0}1\underline{1}0
\underline{0}1\underline{0}1\underline{1}0\cr&\dots}
\eqno(4.5)
$$
The aperiodic chain is constructed through the correspondences
$0\!\rightarrow\!\lambda$ and $1\!\rightarrow\!\lambda r$.
It is clear from (4.4) that the asymptotic
density of modified bonds is $\rho_\infty\!=\!1/2$ and, according
to (3.3),
$$
\lambda_c=r^{-1/2}\eqno(4.6)
$$

{\par\begingroup\parindent=0pt\medskip
\epsfxsize=9truecm
\topinsert
\centerline{\epsfbox{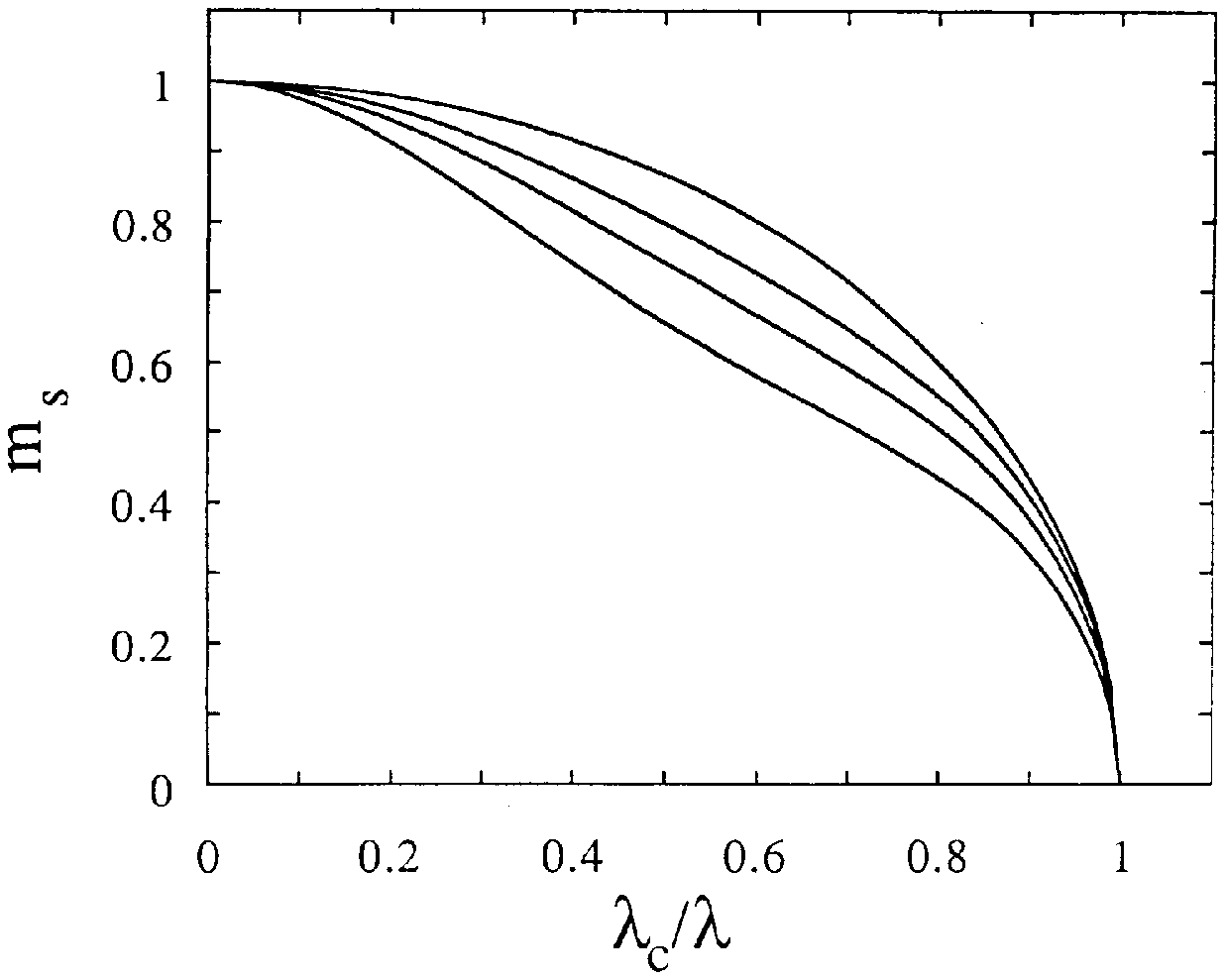}}
\smallskip
\figure{Surface magnetization of the binary Thue-Morse 
aperiodic chain with a modulation amplitude $r=1,2,3,5$ from top
to bottom.}
\endinsert 
\endgroup
\par}
This value of the density is also reached for any finite 
sequence with an even length so that the number
of modified bonds satisfies
$$
n_{2k}=k\eqno(4.7)
$$
The substitution (4.4) is such that the binary Thue-Morse
sequence is reproduced by selecting the sequence of its odd
elements ($f_{2k+1}^{(2)}\!=\!f_{k+1}^{(2)}$) as indicated in the
last line of (4.5). As a consequence
$$
n_{2k+1}=k+f_{k+1}^{(2)}\qquad k=0,1,2\dots\eqno(4.8)
$$
The sum involved in the surface magnetization (2.6) can then be
written as follows
$$
\sum_{j=1}^\infty\prod_{i=1}^j\lambda_i^{-2}=\sum_{k=1}^\infty
\left({\lambda_c\over\lambda}\right)^{4k}+\sum_{k=0}^\infty
\left({\lambda_c\over\lambda}\right)^{2(2k+1)}r^{1-2f_{k+1}^
{(2)}}\eqno(4.9)
$$

{\par\begingroup\parindent=0pt\medskip
\epsfxsize=9truecm
\topinsert
\centerline{\epsfbox{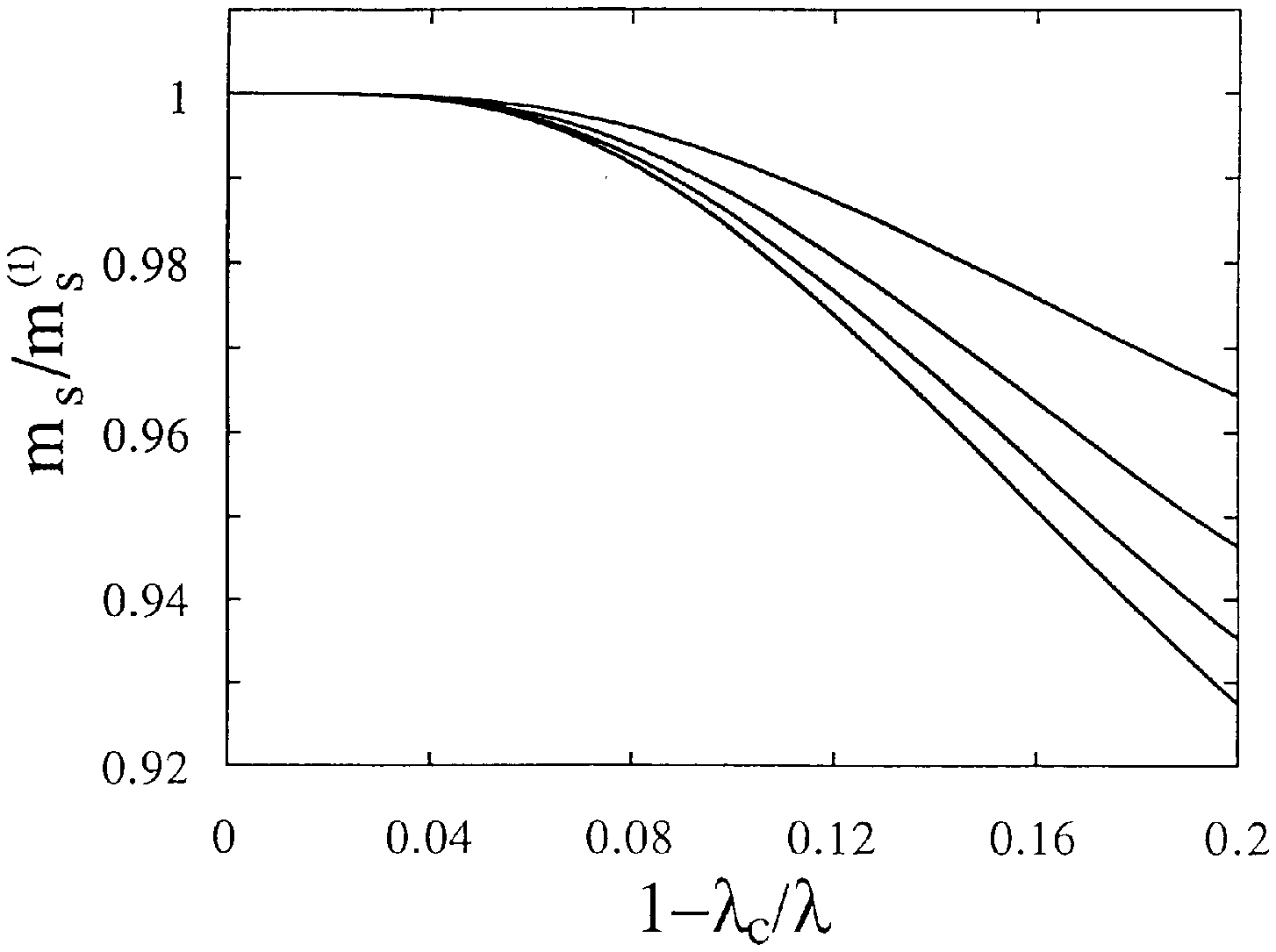}}
\smallskip
\figure{Ratio of the binary Thue-Morse surface magnetization
$m_s$ to its leading critical contribution $m_s^{(1)}$ for a
modulation amplitude $r=2,3,4,5$ from top to bottom. The 
corrections to scaling, due to the higher regularity of the 
sequence, are much weaker than for the Fibonacci sequence.}
\endinsert 
\endgroup
\par}

Near the critical point the last term may be replaced, as above, 
by its average over the infinite sequence in which
$0$ and $1$ occur with the same probability. Then
$$
\left<r^{1-2f_{k+1}^{(2)}}\right>={1\over 2}\left(r+r^{-1}
\right)\eqno(4.10)
$$
and the exact leading critical behaviour reads
$$
m_s^{(1)}={2\sqrt{r}\over r+1}\left[1-\left(\lambda_c
\over\lambda\right)^2\right]^{1/2}\eqno(4.11)
$$
The surface magnetization obtained numerically using (2.6) 
is plotted in figure $3$ for different
values of $r$ and the ratio $m_s/m_s^{(1)}$ is shown in figure $4$.
As a result of the higher regularity of the Thue-Morse sequence, 
the corrections to scaling are much weaker than for
the Fibonacci sequence.
\section{Thue-Morse sequences modulo $p$}
The results of the last section can be generalized by 
considering Thue-Morse sequences $f_j^{(p)}$ defined modulo 
$p$. These sequences are generated by taking the sum modulo 
$p$ of the digits of non-negative integers written in base 
$p$. With $p\!=\!3$, for example, one obtains
$$
0\quad 1\quad 2\quad 10\quad 11\quad 12\quad 20\quad 21\quad 22
\quad 100\quad 101\quad 102\quad\dots\eqno(5.1)
$$
and the sequence $f_j^{(3)}$ reads
$$
012120201120\dots\eqno(5.2)
$$
$f_j^{(p)}$ also follows from the substitution
$$
0\rightarrow 012\dots p\!-\!1\quad 1\rightarrow 12\dots 
p\!-\!1\ 0\quad\dots\quad p\!-\!1\rightarrow p\!-\!1\ 0
\dots p\!-\!2\eqno(5.3)
$$
which for $p\!=\!3$ leads to
$$
\eqalign{&0\cr&012\cr&012120201\cr&\underline{0}12
\underline{1}20\underline{2}01\underline{1}20\underline{2}01
\underline{0}12\underline{2}01\underline{0}12\underline{1}20
\cr&\dots}
\eqno(5.4)
$$
One associates a modified coupling on the chain 
with any nonzero digit along the sequence, \ie\  we use the
correspondence
$$
0\rightarrow\lambda\qquad 1,2,\dots,p\!-\!1
\rightarrow\lambda r\eqno(5.5) 
$$
Then, according to (5.3), the asymptotic density of modified 
bonds is $\rho_\infty\!=\!(p\!-\!1)/p$ 
and the critical coupling $\lambda_c\!=\!r^{-(p\!-\!1)/p}$. 
The asymptotic density is also obtained when the number of bonds
$n_j$ is any multiple $pk$ of $p$ so that 
$$
n_{pk}=pk\rho_\infty=k(p-1)\eqno(5.6)
$$
One may check on the last line of (5.4) that $f_{pk+1}^{(p)}\!
=\!f_{k+1}^{(p)}$ whereas $f_{pk+2}^{(p)},f_{pk+3}^{(p)}\dots$ 
are deduced
from $f_{k+1}^{(p)}$ through circular permutations on
$0,1,2,\dots ,p\!-\!1$. Then, the number of
modified bonds after $pk\!+\!q$ steps can be written as
$$
n_{pk+q}=k(p-1)+m_{k+1}(q)\qquad 0\leq q<p-1\eqno(5.7)
$$
where 
$$
m_k(q)=\sum_{i=0}^{q-1}g\!\!\left[\left(f_k^{(p)}+i
\right)_{\mod\  p}\right]\quad 0<q<p-1,\qquad m_k(0)=0\eqno(5.8)
$$
and
$$
g(i)=1-\delta_{i,0}\eqno(5.9)
$$
The sum in (2.6) may be splitted into $p$ parts and rewritten 
as
$$
\sum_{j=1}^\infty\prod_{i=1}^j\lambda_i^{-2}=\sum_{k=1}^\infty
\left({\lambda_c\over\lambda}\right)^{2pk}+\sum_{q=1}^{p-1}
\left({\lambda_c\over\lambda}\right)^{2q}r^{2q(p-1)/p}
\sum_{k=0}^\infty\left({\lambda_c\over\lambda}\right)^{2pk}
r^{-2m_{k+1}(q)}\eqno(5.10)
$$
When $\lambda\!\rightarrow\!\lambda_c$, $r^{-2m_{k+1}(q)}$ in the 
last sum may be replaced by its averaged value. On 
the infinite sequence, $m_k(q)$ is equal to $q\!-\!1$ with 
probability $q/p$ and equal to $q$ with probability $(p\!-\!q)/p$, 
so that
$$
\left<r^{-2m_{k+1}(q)}\right>={qr^{-2(q-1)}+(p-q)r^{-2q}\over p}
\eqno(5.11)
$$
In the critical region, one finally obtains 
$$
m_s^{(1)}=\sqrt{p}\left({r^{1/p}-r^{-1/p}\over r-r^{-1}}\right)
\left[1-\left(\lambda_c\over\lambda\right)^{2p}\right]^{1/2}
\eqno(5.12)
$$
for the leading contribution to the surface magnetization in
agreement with (4.11) when $p\!=\!2$.
{\par\begingroup\parindent=0pt\medskip
\epsfxsize=9truecm
\topinsert
\centerline{\epsfbox{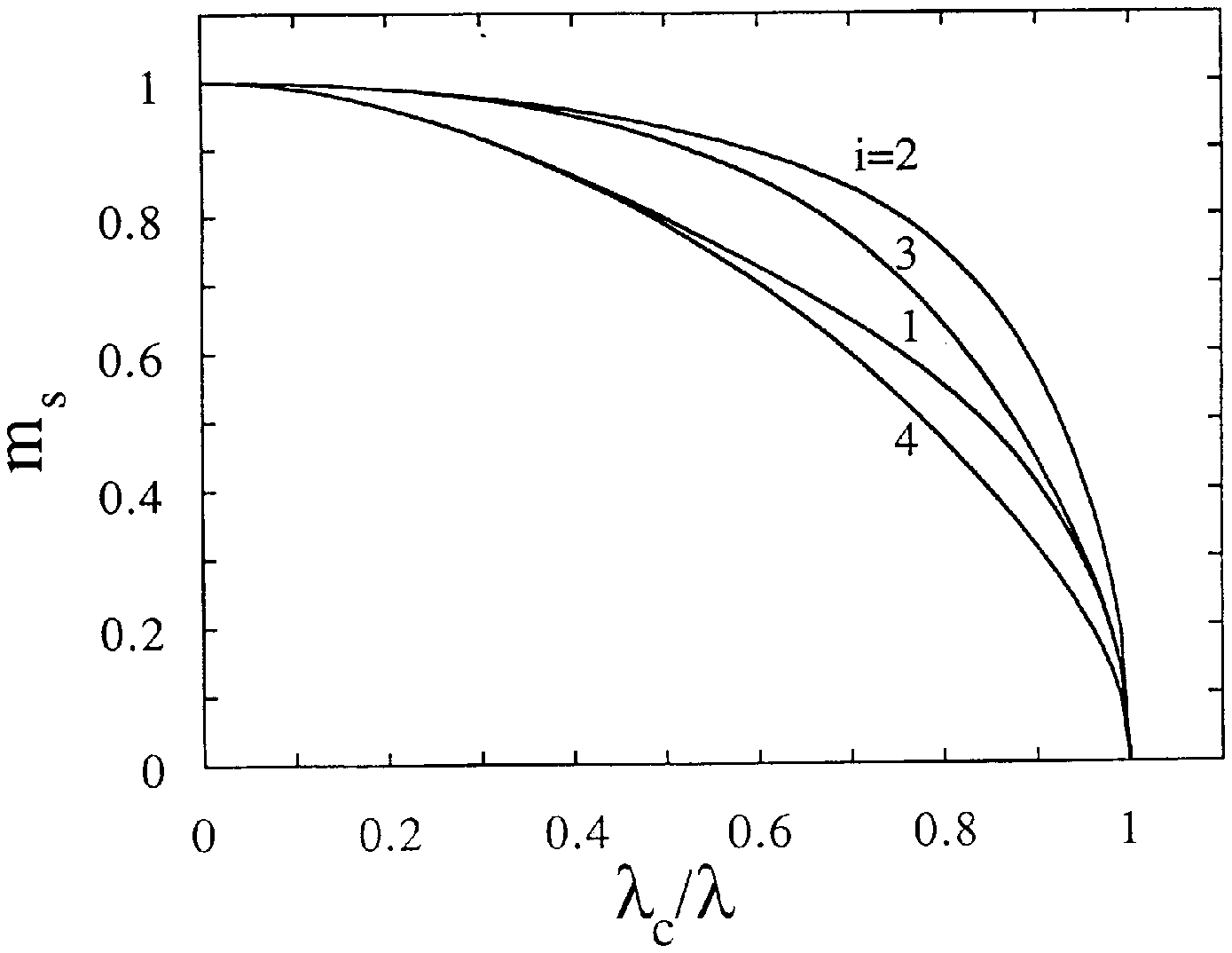}}
\smallskip
\figure{Surface magnetization of binary Thue-Morse aperiodic
chains with $r=2$, starting on the $i$-th digit along the sequence
with $i=1,2,3,4$. The amplitude of the leading term given by
(4.11) when $i$ is odd, is the geometric mean of the amplitudes
obtained for even values of $i$. In the strong coupling regime
the local environment governs the behaviour of the surface 
magnetization.}
\endinsert 
\endgroup
\par}
\section{Discussion}
In the cases studied so far the modulation was distributed 
on the spin chain according to a given sequence, starting with
the first digit of the sequence. One expects a change in the 
surface magnetization amplitude for chains starting elsewhere
along the sequence. The previous calculations are easily
extended to the case where one begins the chains on the 
$i$-th digit.

{\par\begingroup\parindent=0pt\medskip
\epsfxsize=9truecm
\topinsert
\centerline{\epsfbox{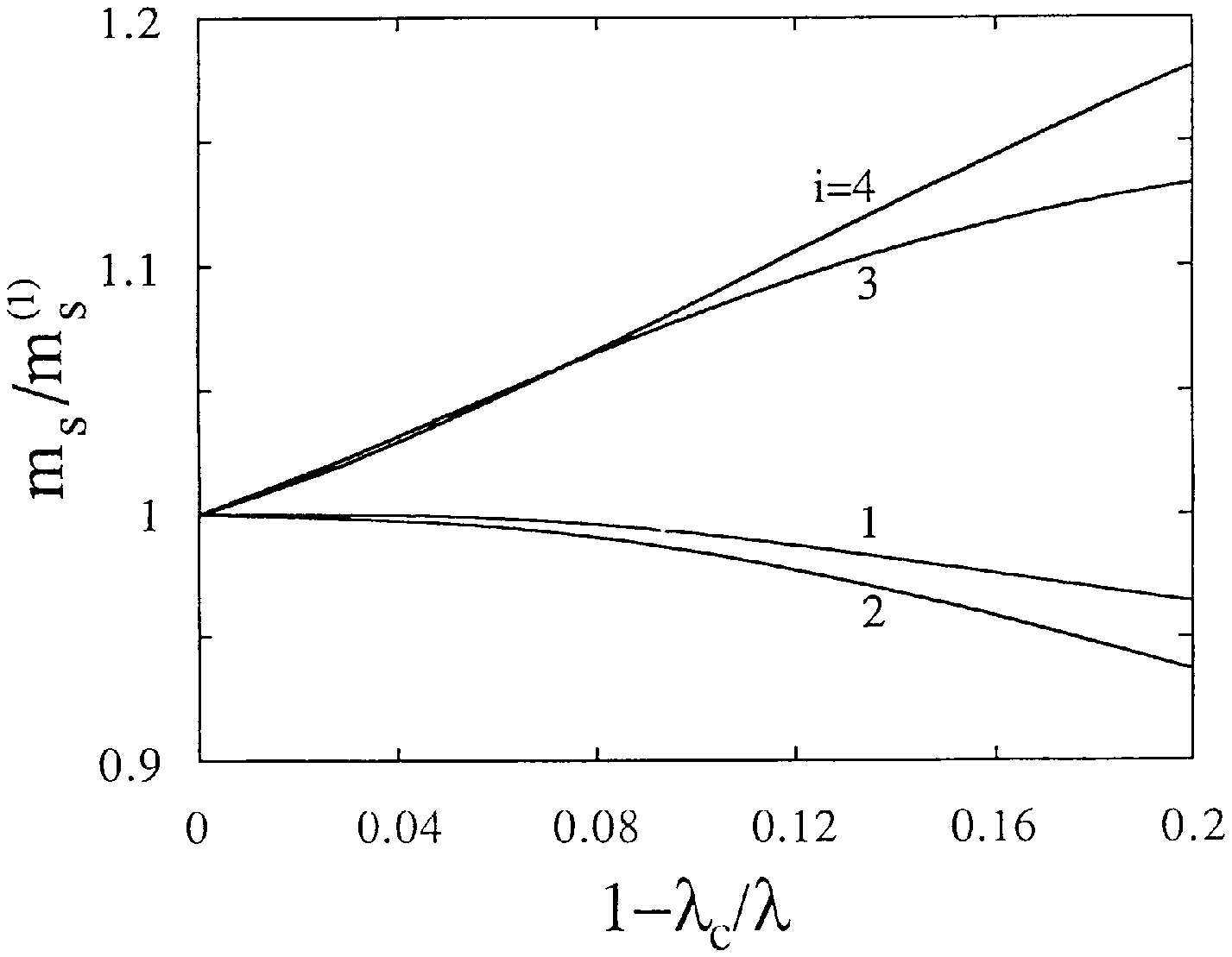}}
\smallskip
\figure{Ratio of the surface magnetization $m_s$ to its 
leading term $m_s^{(1)}$ for the binary Thue-morse Ising chain
with $r=2$ and different starting points $i$ along the sequence.
Although the leading amplitude is the same for all chains with
odd values of $i$, the corrections to scaling are different.}
\endinsert 
\endgroup
\par}

For the Fibonacci sequence, $n_j$ in (3.6) is changed into
$$
n_j=j\left(1-{1\over\omega}\right)+\left\{{j+1\over\omega}
\right\}-\left\{{i\over\omega}\right\}\eqno(6.1)
$$
leading to
$$
m_s^{(1)}(i)=r^{1-\{{i/\omega}\}}\left[{\ln r^2\over r^2-1}\right]^{1/2}
\left[1-\left({\lambda_c\over\lambda}\right)^2\right]^{1/2}
\qquad i>1\eqno(6.2)
$$
With the binary Thue-Morse sequence starting on $i$ even,
\ie with $f_{2l}^{(2)}$, one obtains
$$
n_{2k}=k-1+f_{k+l}^{(2)}+f_{2l}^{(2)}\qquad n_{2k+1}=k+
f_{2l}^{(2)}\qquad i=2l\eqno(6.3)
$$
and then, on the surface site,
$$
m_s^{(1)}(2l)={2r^{f_{2l}^{(2)}}\over r+1}\left[1-\left(
{\lambda_c\over\lambda}\right)^2\right]^{1/2}\eqno(6.4)
$$
Otherwise, when $i$ is odd,
$$
n_{2k}=k\qquad n_{2k+1}=k+f_{k+l}^{(2)}\qquad i=2l+1\eqno(6.5)
$$
like in (4.7-8). Then the leading 
term is still given by (4.11) which is just the geometrical
mean of the two allowed values in (6.4). 
The dependence of the
surface magnetization on $i$ is shown in figure $5$. The local
environment clearly governs the behaviour when $\lambda_c/\lambda
\rightarrow 0$. The surface magnetization is reduced when the
surface site is linked to the bulk via a weak bond for $i=1$ and 
$4$. Near the critical point, on the contrary, only the long-range 
structure of the modulation and the parity of the surface site
are important.

Although the amplitude of the leading contribution to the
surface magnetization is the same for all the aperiodic chains
starting with odd values of $i$ along the Thue-Morse sequence
the corrections to scaling differ as shown on figure $6$.

Similar results would be obtained with the Thue-Morse sequence
defined modulo $p$ with the same leading amplitude for chains 
starting on $i=pl+1$ along the sequence.

For all the sequences considered here, the aperiodic 
modulation, either quasiperiodic or automatic, leaves the 
surface magnetic exponent unaffected,
$\beta_s=1/2$, like in the homogeneous semi-infinite Ising 
model. On the other hand, the critical amplitude depends on 
the modulation and it would be of interest to study other 
surface quantities in order to check whether surface critical
amplitude ratios remain universal. The study of aperiodic 
sequences which do not lead to ordinary bulk Ising behaviour
[9] would also be worthwhile.

\references
\numrefjl{[1]}{Shechtman, D., Blech, I., Gratias, D., Cahn, J.W.}
{\PRL}{53}{(1984)\ 1951}
\numrefjl{[2]}{Henley, C.L.}{Comments Cond. Mat. Phys.}{13}{(1987)\ 59}
\numrefjl{[3]}{Janssen, T.}{Phys. Reports}{168}{(1988)\ 55}
\numrefjl{[4]}{Guyot, P., Kramer, P., de Boissieu, M.}{Rep. 
Prog. Phys.}{54}{(1991)\ 1373}
\numrefjl{[5]}{Janot, C., Dubois, J.M., de Boissieu, M.}{Am. J. 
Phys.}{57}{(1989)\ 972}
\numrefjl{[6]}{Godr\`eche, C., Luck, J.M., Orland, H.J.}{J. Stat.
Phys.}{45}{(1986)\ 777}
\numrefjl{[7]}{Langie, G., Igl\'oi, F.}{J. Phys. A}{25}{(1992)\ L487}
\numrefjl{[8]}{Luck, J.M., Nieuwenhuizen, Th.M.}{Europhys. Lett.}{
2}{(1986)\ 257}
\numrefjl{[9]}{Tracy, C.A.}{J. Phys. A}{21}{(1988)\ L603;\ {\it J. Stat.
Phys.}\ {\bf 51}\ {\rm (1988)\ 481}}
\numrefjl{[10]}{Igl\'oi, F.}{J. Phys. A}{21}{(1988)\ L911}
\numrefjl{[11]}{Ceccatto, H.A.}{\PRL}{62}{(1988)\ 203;\ {\it Z. 
Phys. B}\ {\bf 75}\ {\rm (1989)\ 253}}
\numrefjl{[12]}{Henkel, M., Patk\'os, A.}{J. Phys. A}{25}{(1992)\ 5223}
\numrefjl{[13]}{Bombieri, E., Taylor, J.E.}{J. Physique 
Colloq.\ C3}{47}{(1986)\ 19;\ {\it Contemp. Math.}\ {\bf 64}\ {\rm (1987)\ 241}}
\numrefjl{[14]}{Lieb, E.H., Schultz, T.D., Mattis, D.C.}{Ann. 
Phys. (N.Y.)}{16}{(1961)\ 406}
\numrefjl{[15]}{Peschel, I.}{Phys. Rev. B}{30}{(1984)\ 6783}
\numrefjl{[16]}{Jordan, P., Wigner, E.}{Z. Phys.}{47}{(1928)\ 631}
\numrefjl{[17]}{Pfeuty, P.}{Ann. Phys. (N.Y.)}{57}{(1970)\ 79}
\numrefjl{[18]}{Pfeuty, P.}{Phys. Lett.}{72A}{(1979)\ 245}
\numrefjl{[19]}{Axel, F., Allouche, J.P., Kl\'eman, M., 
Mend\`es-France, M., Peyri\`ere, J.}{J. Physique Colloq.\ C3}{47}{(1986)\ 181}
\numrefjl{[20]}{Dekking, M., Mend\`es-France, M., van der Poorten,
A.}{Math. Intelligencer}{4}{(1983)\ 130}

\vfill\eject\bye